\documentclass[12pt]{iopart}

\usepackage{graphicx}  
\usepackage{subeqnar}  
\usepackage{threeparttable}

\begin{document}


\title[Robustness of the global non-analytical fit to the analysis of
peak-dip-hump lineshape...] {Robustness of the global
non-analytical fit to the analysis of peak-dip-hump lineshape in
bilayer-split spectra of Bi$_2$Sr$_2$CaCu$_2$O$_{8+\delta}$}

\author{R. H. He \footnote[1]{Electronic mail: ruihuahe@stanford.edu;
Present address: Department of Applied Physics, Stanford
University, Stanford, California 94305}}

\address{Physics Department, Applied Surface Physics
State Key Laboratory, Fudan University, Shanghai 200433, China}

\date{\today}

\begin{abstract}
The application of global non-analytical fit to the quantitative
spectral analysis of the bilayer-split peak-dip-hump lineshape of
Bi$_2$Sr$_2$CaCu$_2$O$_{8+\delta}$ at $(\pi,0)$ are revisited. The
robustness of the results is verified by various fitting schemes
confirming physically the correlation of the superconducting peak
with the superfluid density.
\end{abstract}


\maketitle

The most dramatic variation across the superconducting transition
temperature ($T_c$) in spectral function is the development of
\textit{peak-dip-hump} (PDH) structure near
$(\pi,0)$\cite{ZXReview}. This has been observed by angle-resolved
photoemission spectroscopy (ARPES) in many high temperature
superconductors (HTSC's), and is thought to carry critical
information about the superconducting
transition\cite{ubiquitousPDH}. Particularly, on
Bi$_2$Sr$_2$CaCu$_2$O$_{8+\delta}$ (Bi2212), we found the
superconducting peak (SCP) intensity to be closely related to the
condensate fraction of the system\cite{FengScience}. This was
achieved by a spectral analysis based on the SCP ratio (SPR)
accounting for the phenomenologically-extracted coherent component
in $(\pi,0)$ ARPES lineshape. The phase-coherent like behavior of
SPR (increases with doping increases) in the underdoped regime was
later confirmed by H. Ding et al. based on a similar analysis of
the lineshape by means of the so-called renormalized quasiparticle
weight (RQW), a quantitive equivalent to SPR\cite{DingPRL01}.
However, as a contrast, the decrease of the SPR upon doping
increase across the optimal doping was not reproduced. This led
the authors to think it was necessary to combine the additional
effect by the gap opening to yield, rather than simply taking the
SPR as, the long-sought supercoducting order parameter for the
cuprates.

This controversy remains until later another important finding was
made, the existence of bilayer band splitting (BBS) in heavily
overdoped (OD) Bi2212\cite{FengPRL01}. This gives birth to a sharp
anti-bonding band remnant (AB hump) even in the normal state at
the SCP binding energy position tending to invalidate the SPR
(RQW)-based phenominological background substraction: the SPR
(RQW) obtained this way contains a substantial spectral weight
contribution from the AB hump, which should not be counted in in
order to obtain a true ratio reflecting the SCP only. The
increasing overestimation of the true peak ratio prevents a
similar analysis from applying to the more OD regime where the AB
band becomes sharper. In Ref. \cite{MyRapComm}, a lineshape
modelling based on a physically sensible description of the two
quasi-particle bands\cite{Fink2} proved its eligibilty to
reproduce the complex PDH lineshape under a considerable BBS
complication and yielded a reduced SPR strengthening its
correlation with the superfluid density in the very OD regime.

In this paper, we revisit the investigation by presenting more
results reflecting the robustness of the conclusion, which include
the insensitiveness to a physical variation in the spectral
function, to a different self-normalization scheme adopted.

\begin{figure}
\centerline{\includegraphics[width=5.5in]{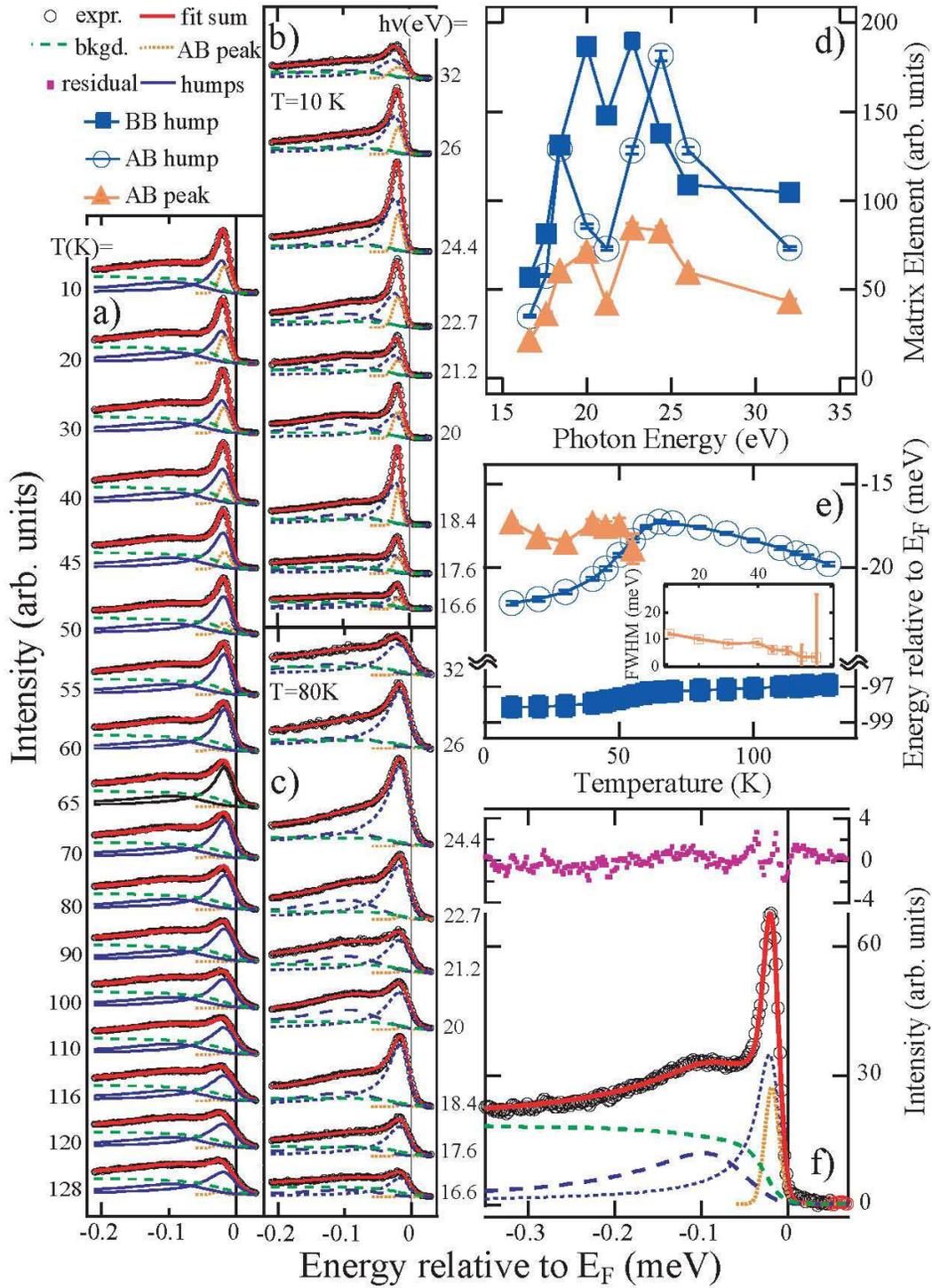}}
\caption[Results of SchemeII] {(color) Results of Scheme II of the
global convolve fit on the $T$-dependent (a) (see also [f] for an
enlargement of the $T$=30 K curves) and $h\nu$-dependent ARPES
spectra at $(\pi,0)$ of (Pb)-OD65 in the superconducting state at
$T$=10 K (b) and normal state at $T$=80 K (c). The AB hump lies
lower in energy than the BB hump. The width difference of the SCP
is due to resolution variations from 10 to 18 meV at different
$h\nu$'s. The fitted $h\nu$-dependent matrix elements (d) and
$T$-dependent energy positions of the maxima (e) of various
spectral features are shown reminiscent of the similar results in
Ref. \cite{MyRapComm}. Inset of (e) illustrates the constancy of
the AB peak width upon $T$ variation in the superconducting
state.} \label{fig2}
\end{figure}

For clarity, we recap below the fitting function used for the
lineshape modelling in Refs. \cite{MyRapComm,MyAlgorithm}. Note
that only possible variations in some expressions are pointed out
and the denominations of symbols completely follow those in Refs.
\cite{MyRapComm,MyAlgorithm} otherwise.
\begin{eqnarray}
\label{eq:Isum}
I(\omega,T,h\nu)&=&I_0(T,h\nu)\cdot[(\sum_{\alpha}^4J_{\alpha}(\omega,T,h\nu)\cdot f(\omega,T)) \nonumber \\
&&\otimes R(\omega,\Gamma^\prime(h\nu))+B(\omega,T)]+I_1(T,h\nu), \nonumber \\
J_{ah}(\omega,T,h\nu)&=&M_{ah}(h\nu)\cdot C_a(T)\cdot A_h(\omega,T,\alpha,\varepsilon_{ah}), \nonumber \\
J_{bh}(\omega,T,h\nu)&=&M_{bh}(h\nu)\cdot C_b(T)\cdot A_h(\omega,T,\alpha,\varepsilon_{bh}), \nonumber \\
J_{ap}(\omega,T,h\nu)&=&M_{ap}(h\nu)\cdot (1-C_a(T))\cdot A_p(\omega,\Gamma_a(T),\varepsilon_{ap}), \nonumber \\
J_{bp}(\omega,T,h\nu)&=&M_{bp}(h\nu)\cdot (1-C_b(T))\cdot
A_p(\omega,\Gamma_b(T),\varepsilon_{bp}), \nonumber \\
A_h(\omega,T,\alpha,\omega_0)&=&\frac{\xi|\Sigma^{\prime\prime}(\omega,T)|}
    {(\omega-\sqrt{\omega_0^2+\Delta_{sc}(T)^2})^2+\Sigma^{\prime\prime}(\omega,T)^2},\nonumber \\
A_p(\omega,\Gamma(T),\omega_0)&=&\frac{2\sqrt{\ln2}}{\sqrt{\pi}\Gamma(T)}
    \exp[-(\frac{\omega-\omega_0}{\Gamma(T)/2})^2].\nonumber
\end{eqnarray}
, where the imaginary part of the self-energy is defined either by
\begin{eqnarray}
\label{eq:SigmaPrimePrime1}
\Sigma^{\prime\prime}(\omega,T)=\sqrt{(\alpha\omega)^2+(\beta
T)^2}
\end{eqnarray}
, as in Ref. \cite{MyRapComm}, or by
\begin{eqnarray}
\label{eq:SigmaPrimePrime2}
\Sigma^{\prime\prime}(\omega,T)=|\sqrt{(\alpha\omega)^2+(\beta
T)^2}+\zeta|
\end{eqnarray}
\cite{MyAlgorithm}, where lifetime broadening associated with the
impurity level of samples is reflected by $\zeta$ and that with
doping level of by $\alpha$ and $\beta$, which are photon energy
($h\nu$) and temperature ($T$) independent. $\Gamma(T)$ is a
$T$-dependent linewidth (FWHM) for the SCP. Note that, as a sum
rule for an overall spectral weight conservation upon $T$
variation, we adopt either
\begin{eqnarray}
\label{eq:SumRule1}
\int_{-\infty}^{\infty}A_{h(p)}(\omega)d\omega=1
\end{eqnarray}
Refs. \cite{MyRapComm,MyAlgorithm} or
\begin{eqnarray}
\label{eq:SumRule2}
\int_{-\infty}^{\infty}f(\omega)A_{h(p)}(\omega,T)d\omega=n(T)
\end{eqnarray}
\cite{ZXReview}, where $n(T)$ is the occupation number at
$(\pi,0)$, which is nearly independent of $T$\cite{RanderiaPRL95}
and thus assumed to be a constant here. As a result, spectral
weight transfer (SWT) ($=1-C_{a(b)}, 0\leq C_{a(b)}(T)\leq 1$) is
well defined for the description of the SCP formation disregarding
the kind of dependence under investigation\cite{MyRapComm}. In any
case, a realistic SCG or pseudogap opening extracted with the
leading edge gap method on the same sample introduces an effective
$T$-dependent energy shift to the bilayer-split bands in the
superconducting state while the energy position of the SCP is
subjected to the fit.

In order to retrieve reliable quantitative information from a
multi-dimensional fit to a set of interconnected EDC's, we
developed a powerful fitting routine under WaveMetrics Igor Pro
4.0\cite{IgorNote}, which has been applied in our previous
work\cite{MyRapComm} and detailed in Ref. \cite{MyAlgorithm}.
Physically, the goal is to achieve a systematic understanding of
both the $h\nu$- and the $T$-dependent behavior of the lineshape
where many physical constraints are involved and strictly
followed. For example, in the the $h\nu$-dependent set, all
functions of $T$ are shared variants to all EDC's at the same $T$
while the functions of $h\nu$ (matrix elements) are locally
specified, and in the $T$-dependent set, vice versa. The
renormalized band energy position $\varepsilon$ is always a global
fitting parameter for the EDC's of samples at the same doping
level. Technically, reminiscent of the great contrast in the
overall quality (or details) of the fit between the one shown in
Fig. 2 of Ref. \cite{Fink2} and ours, in Refs.
\cite{MyRapComm,MyAlgorithm}, though on data sets generally twice
larger in size, the inherently local fit scheme\cite{Fink2, Fink4}
with manual adjustments on the globally defined fitting parameters
is unlikely to give a reliable result in a truly global sense. In
contrast, our fully-automatic fitting routine can yield a robust,
physically-constrained global fit for a large set of EDC's
regardless of the initial values input.

The global non-analytical fit implemented in Ref. \cite{MyRapComm}
is based on Eq. \ref{eq:SigmaPrimePrime1} and \ref{eq:SumRule1}
(Scheme I). In the following, we conducted another two schemes,
the one with Eq. \ref{eq:SigmaPrimePrime2} and \ref{eq:SumRule1}
(Scheme II) and the other with Eq. \ref{eq:SigmaPrimePrime2} and
\ref{eq:SumRule2} to complement the results of analysis based on
the global non-analytical fit. Our argument about the
qualification of the SPR as the order parameter for the HTSC's is
invulnerable to the physical variation of the fitting scheme we
choose. This checkup is hardly practical without a fully-automatic
multi-level global non-analytical fit routine which can be easily
oriented for any non-analytical global fitting task given.

\begin{figure}[t]
\centerline{\includegraphics[width=5.5in]{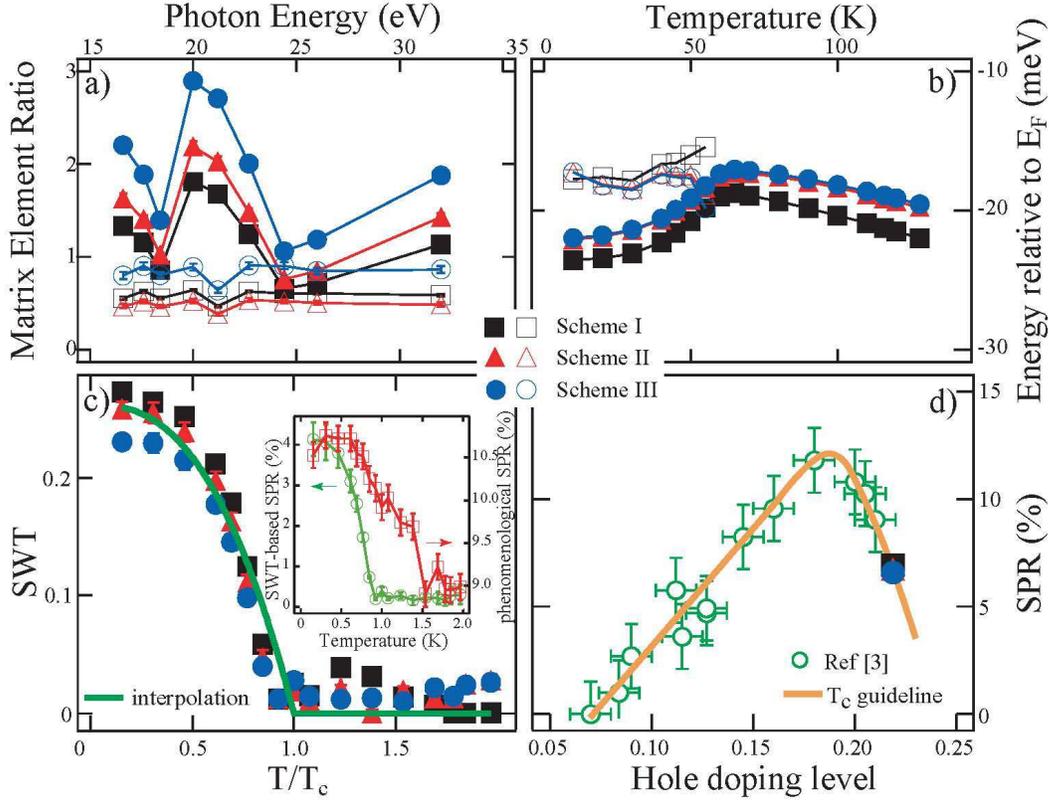}}
\caption[Scheme contrast] {(color) Results of the three schemes of
the global non-analytical fit. The fitted $h\nu$-dependent matrix
element hump ratios, $\frac{M_{bh}}{M_{ah}}$ (filled markers), and
peak-hump ratios, $\frac{M_{ap}}{(M_{ah}+M_{bh})/2}$ (empty
markers) are shown in (a). (b) shows the fitted $T$-dependent
energy positions of the AB peak (empty markers) and AB hump maxima
(filled markers). (c) The $T$-dependent SWT's and an interpolation
on the averaged dependence according to the function
SWT=SWT$_0[1-(T/T_c)^{\zeta}]$ (see Ref. \cite{TallonTl2201}). The
fitted value, $\zeta=2.56$, agrees with the $T$-dependent $\mu$SR
result on $Tl_2Ba_2CuO_{6+\delta}$ at the same doping level (OD66
Tl2201)(not shown). The contrast between the phenomenological SPR
and SWT-based SPR's extracted from the same set of data (taken
with He-I) is given in the inset. (d) The doping-dependence of SPR
($h\nu$=22.7 eV, $T=10$ K) supplemented by the global fit results
on Pb-OD65.} \label{fig3}
\end{figure}

In Fig. \ref{fig2}, we present the results of Scheme II,
reminiscent of Fig. 2-3 in Ref. \cite{MyRapComm}. The fitting
quality is exemplified in Fig. \ref{fig2}(f). Keep in mind that
this quality is common to the fits of the 35 EDC's in contrast to
a local best fit on a few selected curves at the expense of a
global agreement on a much larger assembly within the usual manual
fitting framework\cite{Fink2}. In Scheme II, when we further
release the constrain on $\Gamma_a$ to let it fittable, we find
its $T$-dependence is negligible as obtained by our
phenomenological fitting. This argues against the Fermi liquid
approach that attributes the disappearance of the sharp peak above
$T_c$ to broadening caused by phase
fluctuations\cite{FengScience19,SCPwidth}.

Note that the addition of an impurity term (whose fitted value is
$-0.0058\pm0.0004$ for Pb-OD65 and $-0.0067\pm0.0004$ for OD65) in
the imaginary self-energy doesn't result in any qualitative change
in the $h\nu$-dependence of matrix elements as well as the
$T$-dependence of the energy positions of spectral features [Fig.
\ref{fig2}(d)(e)]. However, quantitative changes is discernible
when we plot the results of the three schemes together as in Fig.
\ref{fig3}. Though the modification of the normalization equation
is not expected to alter much the energy positions of features
whose maximums are more or less self-pinned [Fig. \ref{fig3}(b)]
and the anti-phase-like $h\nu$-dependence of the hump matrix
elements as predicted by theory\cite{LindroosPRB,FengPRB02}, the
matrix element ratio of the two hump bands
($\frac{M_{bh}}{M_{ah}}$) is, in principle, a touchstone for the
justification of the physical normalization method. The average of
matrix element hump ratio over a large $h\nu$ window is supposed
to approach unity as expected from a pair of bands of the same
atomic character split by the $c$-axis bilayer coupling. Though
the $h\nu$ window concerned here is not large enough, Scheme I and
II give a ratio of $0.91\pm0.01$ and $1.21\pm0.04$, respectively,
quite approaching unity, while Scheme III yields $1.55\pm0.02$.

Interestingly, compared with the hump ratio, the matrix element
ratio between the peak and the hump, defined by
$\frac{M_{ap}}{(M_{ah}+M_{bh})/2}$, exhibits a constancy over
$h\nu$. Reminiscent of the absence of SCP finding report in
spectra of mono-layer cuprate Bi$_2$Sr$_2$CuO$_{6+\delta}$, this
probably imply the AB peak has a hybrid origin of AB and BB humps
at the same time. It is argued by the magnetic resonance
($\pi,\pi,\pi$) mode scenario that the c-axis momentum transfer by
the mode scattering occurs between AB and BB which leads to the
formation of the PDH lineshape\cite{Campuzano1}. However, in our
picture the SCP is a new quasi-particle developed from its normal
state counterpart of hump, just the opposite to the magnetic mode
scenario where the dip(hump) is argued to contain the incoherent
spectral weight subjected to the scattering of the quasi-particle
peak. This critical question concerning the SC origin is still
unaddressed.

As told by Fig. \ref{fig3}(c), the SWT saturates at low
temperature to be ($0.26\pm0.03$), a value far below 1, which is
required by the magnetic mode scenario in which the collective
mode induces an on/off damping mechanism at the dip position
driving the whole lower-lying AB component coherent with no AB
hump remanence\cite{NormanPRL02}. As mentioned before, the large
remanence of AB hump proportion at the experimentally observed SCP
energy position in the superconducting state also questions the
validation of the SPR (RQW)-based phenomenological background
substraction\cite{FengScience,DingPRL01} in the very OD regime.
This can be illustrated by the contrast in the inset of Fig.
\ref{fig3}(c) with the phenomenological SPR featured by an overall
overestimation in value and a non-zero normal state baseline
followed by a rounded onset spreading over a large $T$ region
across $T_c$.

The collapses of the $T$-dependent SWT (SPR) onto the superfluid
density curve [Fig. \ref{fig3}(c)] and of the SWT-based SPR's by
the three schemes onto the $T_c$ guideline strength [Fig.
\ref{fig3}(d)] strengthen again the qualification of SPR (SWT) as
a superconducting order parameter for Bi2212. However, the
component-resolved matrix-element-free merit of SWT makes it an
ideal quantity for the quantitative spectral analysis of the
complex lineshape at $(\pi,0)$. A systematic investigation of both
the doping and $T$ dependence of SWT based on an indispensable
global fit facility is needed for a comprehensive understanding of
its relation to the possible mechanism for high-temperature
superconductivity.

I thank Dr. D. L. Feng, D. H. Lu, and W. S. Lee at SSRL for
experimental help, K. M. Shen and Dr. A. Damacelli for comments
and discussions. SSRL is operated by the DOE Office of Basic
Energy Science Divisions of Material Sciences.




\begin{thebibliography}{}
\bibitem{IgorNote} Igor uses the Levenberg-Marquardt algorithm to search for the
coefficient values that minimize $\chi^2$. This is a form of
non-linear, least-squares fitting. This kernel is well-packed and
convenient for customizing interfaces oriented for different uses.
\end{thebibliography}
\end{document}